\documentclass[a4paper]{jpconf}
\usepackage{amssymb}
\usepackage{graphicx}
\usepackage{amsthm}
\usepackage{latexsym}
\newcommand\beq{\begin{equation}}
\newcommand\eeq{\end{equation}}
\newcommand\bea{\begin{eqnarray}}
\newcommand\eea {\end{eqnarray}}

\def\a{{\alpha}}

\def\Tr{{\mathrm{Tr}}}

\begin{document}
\title{\textbf{The coevent formulation of quantum theory}}

\author{\textbf{Petros Wallden}}
\address{1. University of Athens, Physics Department, Nuclear \& Particle Physics Section, Panepistimiopolis 157-71, Ilissia
Athens, Greece}

\address{2. Technological Educational Institute of Chalkida, General Science Department,  Psahna-34400, Greece}

\ead{\textbf{petros.wallden@gmail.com, pwallden@phys.uoa.gr}}

\date{}

 \vspace{1cm}

\begin{abstract}

Understanding quantum theory has been a subject of debate from its birth. Many different formulations and interpretations have been proposed. Here we examine a recent novel formulation, namely the coevents formulation. It is a histories formulation and has as starting point the Feynman path integral and the decoherence functional. The new ontology turns out to be that of a coarse-grained history. We start with a quantum measure defined on the space of histories, and the existence of zero covers rules out single-history as potential reality (the Kochen Specker theorem casted in histories form is a special case of a zero cover). We see that allowing coarse-grained histories as potential realities avoids the previous paradoxes, maintains deductive non-contextual logic (alas non-Boolean) and gives rise to a unique classical domain. Moreover, we can recover the probabilistic predictions of quantum theory with the use of the Cournot's principle. This formulation, being both a realist formulation and based on histories, is well suited conceptually for the purposes of quantum gravity and cosmology.

\end{abstract}

\section{Motivation}

Quantum theory is undoubtedly one of the most successful theories. The understanding and interpretation of quantum theory, has been a subject of debate from the birth of the theory until now. There is no unique, widely accepted interpretation of quantum theory. In the recent years, due to the developments in the field of quantum information and the search of a quantum theory of gravity and cosmology, there is a new wave of interest for the foundations of quantum theory. This contribution, is a review of a novel formulation of quantum theory namely the ``coevents formulation''. The motivation for such a formulation, is twofold.

First is the need for a realist interpretation of quantum theory. The standard view of quantum theory, is that of an instrumentalist. Avoids to refer to the actual ontology, and makes assertions only related with some idealised measurements carried out by an external observer. However, particularly for the field of quantum cosmology where one treats the full universe as a quantum system, the need for an interpretation that does not require an external observer becomes vital. In this sense the need for a realist interpretation, becomes more than a philosophical investigation.

Second is the need for a formulation that treats space and time in equal footing. In constructing a quantum theory of gravity, one is led to the observation that while space and time are totally different objects in quantum theory, they are more-or-less the same in general relativity. This leads to a tension that has both technical and philosophical difficulties (e.g. see the problem of time \cite{problem of time}). We can claim that better suited for the purpose of quantum gravity should be a formulation of quantum theory that is based on full histories of the system (such as in the Feynman path integral) and \emph{not} a canonical formulation (that relies on the Hamiltonian and uses one-moment-of-time propositions). The latter may not even be applicable for certain approaches to quantum gravity such as the causal sets.

The coevents formulation was introduced by Rafael Sorkin in \cite{coevent1} following his earlier work in \cite{coevent2} (see section \ref{3.0}). It is a realist interpretation, that is based on histories. In particular, the starting point is the Feynman amplitudes for histories, and interpreting those amplitudes in an observer-free, context-independent way is the aim of the formulation. In section \ref{2.0} we will give the standard histories view for classical physics, state which is the importance of precluded sets and stress which is the problem in adopting a ``naive'' realist view in quantum physics. In section \ref{3.0} we will introduce the coevents formulation, giving two different (but equivalent) views. In section \ref{4.0} we will explore (briefly) certain developments of the formulation. In particular we will define what a classical domain is, show that there exist a unique, context-independent such domain and point out that this is not the case for the consistent histories approach\footnote{The approach is also known as \emph{decoherent histories}.} (section \ref{4.1}). We will give an account of how probabilistic predictions arise (section \ref{4.2}), and point out the role and the status of the initial state in the formulation (section \ref{4.3}). In section \ref{5.0} we will summarise and conclude.

\section{Histories, classical physics and quantum measure}\label{2.0}

We are taking a histories view. Here we will review classical physics casted in histories perspective and stressing certain important features that become important in quantum physics.

\subsection{General Structure}\label{2.1}

In describing physics with histories, there are three important mathematical structures.

\noindent\textbf{(1)} The first, is the space of
all finest grained descriptions $\Omega$. In probability theory, it is called sample space, while in histories
formulation, is called \emph{history space}. The histories space, in classical physics is
the space of potential realities. However this picture does not carry over in quantum theory, as we will see, where potential realities are no longer fine-grained descriptions.

Each element of this space $h\in\Omega$, corresponds to a full description of the system,
specifying every detail and property. For example, a fine grained
history gives the exact position of the system along with the
specification of any internal degree of freedom, for every moment of
time. For a single non-relativistic particle, $\Omega$ would be the space of all trajectories in the physical space.

Along with $\Omega$, we define the space $\mathcal U$, which
consists of all the subsets $A\subseteq\Omega$ and the
Boolean algebra associated with them (called events algebra), where
the addition is defined as the symmetric difference between subsets
$A+B:=A\bigtriangleup B$ and the
multiplication given by the intersection of subsets $A\cdot B:=A\cap
B$. Each subset $A\subseteq\Omega$ is called \emph{event}, as in probability theory. Note, that all
physical questions that one can ask, correspond to one of those
subsets. If, for example, one wishes to ask ``was the system at the
region $\Delta$ at time $ t$?'', it corresponds to the subset
$A$ defined as $\{A: h_i\in A \textrm{ iff } h_i(t)\in\Delta\}$,
i.e. all histories that the system at time $t$ is in the region $\Delta$.

\noindent\textbf{(2)} The second structure, is the space of
truth values (for which we will use the notation $\mathcal T$), and
the algebra associated with them. In the case of classical physics,
the truth values are simply the two elements set
$\mathcal{T}:=\{\textrm{True, False}\}$ (or simply $\{1,0\}$), and
the Boolean algebra, called the truth values algebra,
associated with them. We have $1+1:=0, 0+0:=0, 1+0:=1, 0\cdot 1:=0, 1\cdot
1:=1,0\cdot0:=0$. It is possible to use as truth
values a more general algebra.

\noindent\textbf{(3)} The third structure, is the space of
possible \emph{truth valuations maps} ($\phi:\mathcal U\rightarrow
\mathcal T $). These maps assign a truth value,
to each of the possible questions (events). We shall use the
notation $\mathcal M$ for this space. However, to be able to reason, we need to be able to make deductions, in other words to be able to obtain the truth value of some event from the truth values of some other events. The most strict condition, that holds in classical physics, is that the allowed truth
valuation maps $\phi$ are homomorphisms between the
events algebra $\mathcal U$ and the truth values algebra $\mathcal
T$. In other words we require the map $\phi$ to obey:

(a) Multiplicativity \beq\label{multiplicativity}\phi(A\cdot
B)=\phi(A\cap B)=\phi(A)\phi(B)\eeq

(b) Additivity
(Linearity)\beq\label{additivity}\phi(A+B)=\phi(A\triangle
B)=\phi(A)+\phi(B)\eeq

An important observation, is that maps $\phi$ that are homomorphisms\footnote{In the rest paper, when we refer to homomorphisms or homomorphic maps, it should be understood as homomorphisms between the event algebra $\mathcal U$ and the truth values algebra $\mathcal T$.}, are in a one-to-one correspondence with
single elements $h$ of the space of histories $\Omega$. In particular, maps that are the characteristic function for a particular history $h$, are homomorphisms between the Boolean event algebra and the
Boolean truth values algebra.

\bea \phi_h(A)&=& 1\textrm{ if } h\in A\nonumber\\\phi_h(A)&=&
0\textrm{ otherwise } \eea
Moreover, \emph{all} homomorphisms are of this type. There exists one homomorphic map for each one of the (single) elements $h$ of $\Omega$. Due to the one-to-one correspondence of homomorphic maps and single elements $h$, we could adopt a dual view, and state that classical reality is a homomorphic map $\phi_h$
between the event algebra $\mathcal U$ and the truth values algebra
$\mathcal T$. To sum up, for classical physics, potential realties can be viewed either as single histories $h$ elements of $\Omega$ or as homomorphic maps $\phi_h$. The potential realities are further restricted by the dynamics. As we will see later in section \ref{3.0}, the ontology in quantum physics changes, but the existence of a dual picture of reality as an event or as a map is maintained.

We have explored, so far, the ``kinematic'' part of the theory. We have not mentioned the dynamics (Hamiltonian), or initial condition. In classical deterministic physics, given the initial state and the dynamics, we know all the evolution of the system, and thus the full history. In other words, we know which history $h$ from the space of histories $\Omega$ is actually realised. Given the initial state and dynamics, there is only one possible potential reality $h$, the one that is actually realised. This is the meaning of determinism. Obtaining predictions, becomes trivial, since an event is true if and only if it contains the one realised history. We can define a classical measure $\mu_c$ on $\mathcal U$, such that $\mu_{c}(A)=1\textrm{ if }h\in A$ and $\mu_c(A)=0\textrm{ otherwise}$. This gives the probability that an event $A$ occurs, which is always either one or zero. This simple measure, coincides with the truth valuation $\phi_h$, of the single history that is realised, but should \emph{not} be confused, since the analogy holds only for deterministic classical physics.

In classical stochastic physics, the picture is different. We are not given which history is actually realised, but given the initial state and dynamics, we can obtain a classical measure $\mu_c$ (non-trivial this time) on the space $\mathcal U$ of subsets of $\Omega$. The measure of each event, corresponds to the probability that this event occurs. The measure is no longer related with the valuations $\phi\in\mathcal M$. It is important to note here, that the actual realised reality in stochastic physics, will still be one fine grained history $h$ element of $\Omega$. Since our theory is no longer deterministic, the space of potential such histories has many elements and in particular (for finite histories space) all fine grained histories $h_i$ that have non-zero measure, $\mu_c(h_i)\neq 0$. The role of the measurement in stochastic classical physics, is restricting further the potential realities. For example, before tossing a coin, two outcome were possible, but after performing the measurement and looking the outcome, it is determined wether the outcome was heads or tails. In histories formulation, the latter corresponds to the experimenter restricting the set of possible histories in the universe to those that are compatible with his new observation. To sum up, in stochastic classical physics, the ontology remains the same but the mechanism to obtain predictions changes. We will see more on probabilistic predictions for closed classical or quantum systems, in section \ref{4.2}  after having introduced the coevents formulation.

\subsection{Precluded sets}\label{2.2}

As we have seen, the set of potential realities, both in deterministic and stochastic classical physics, is determined by the classical measure that is defined on $\Omega$, which in its term is fixed by the initial state and dynamics. More specifically, for finite $\Omega$, it is the measure zero sets that determine which are the allowed potential realities/histories \footnote{For infinite $\Omega$ more care is needed, and definition of measurable sets is required. However, most of the arguments carry over.}.  We define \emph{precluded event} to be an event $P\subset \Omega$ such that its measure is zero $\mu_c(P)=0$. The precluded events do not occur. If an event $P$ is precluded, any subset $P_1$ of $P$, should also be ruled out. The latter is a vital condition that needs to be satisfied, if one wishes to make any deductive argument\footnote{This is related with the modus ponens rule of inference. E.g. ``if all humans have two hands'' and ``Plato is human'', we wish to be able to deduce that ``Plato has two hands''.}. In classical physics, this condition is guaranteed by the fact that we require the maps/elements of $\mathcal M$, to be homomorphisms of the Boolean algebras. Simply requiring the condition $\mu_c(P)=0 \Rightarrow \phi(P)=0$ leads directly to $\phi(P_1)=0\textrm{ if } P_1\subseteq P\textrm{ and }\phi(P)=0$. While this discussion seems trivial for classical physics, it will become apparent that it is important for quantum physics.

A final interesting remark, concerning precluded sets, is that one can fully reconstruct all probabilities using only the set of precluded events provided that he has $n\rightarrow\infty$ identical copies of the system. Heuristically, any distribution of outcomes of the identical copies, that differs from the one given by probability, has small chance of occurring, and as the number of copies tends to infinity, this chance of occurring also tends to zero. Technically, let $A$ an event and $A_n$ independent copies. There exists a unique number $p$ such that the following event to be precluded ($I_A$ is indicator function)

\beq\{s\in\Omega^*: (I_A(s_1)+I_A(s_2)+\cdots+I_A(s_n))/n\textrm{ does not tend to } p\textrm{ when } n\rightarrow\infty\}\eeq
See for example \cite{preclusion-probability}. In this sense, all the content of relative frequency interpretation of probability, is included in the precluded sets. Moreover, the precluded sets can be used to recover some predictions via the Cournot principle (see later), even in the absence of many identically prepared copies. This is one more reason why, we choose to give specific importance to precluded events.

\subsection{Quantum measure}\label{2.3}

The picture described above, cannot be (fully) carried over to quantum theory. The histories space $\Omega$ and its subsets/events $\mathcal U$ remains the same. The main difference arises, mathematically, from the fact that we no longer have a classical measure on $\mathcal U$ but rather a \emph{quantum measure}, which we will shortly define. Given an initial condition and dynamics the quantum measure is totally fixed. The initial condition and the dynamics can be given either in form of some initial condition on a path integral along with an action $S$, or as an initial wavefunction in a Hilbert space along with a Hamiltonian operator.

To define the quantum measure (which was first done by Sorkin in \cite{quantum measure}) we need to introduce amplitudes to histories. Starting from Feynman's path integral approach, one can assign an amplitude (complex number) to each history.

\beq\label{Feynman} \a(h_i)=\exp i S(h_i)\eeq
which depends on the initial state and on the dynamics of the system encoded in the action $S$. Obtaining the transition amplitudes from $(x_1,t_1)$ to $(x_2,t_2)$, is done by summing through all the paths $\mathcal P$ obeying the initial and final condition, i.e.

\beq \a(x_1,t_1;x_2,t_2)=\int_\mathcal P \exp(iS[x(t)])\mathcal D x(t)\eeq
The mod square of this amplitude is the transition probability. One can extend this to any event $A\subseteq\Omega$ and proceed to define a quantum measure $\mu$:

\beq \mu(A)=\int_{x\in A}\int_{x'\in A} \exp(iS[x(t)])\exp(-iS[x'(t)])\rho(x(t_0),x'(t_0))\delta(x(t_f)-x'(t_f))\mathcal D x(t)\mathcal D x'(t)\eeq
where the initial state $\rho$ and a final time $\delta$-function is added.

Alternatively, one can take a Hamiltonian view, and obtain the quantum measure in the following way. Define the class operator $C_A$

\beq C_A= P_{A_n}U(t_n-t_{n-1})\cdots U(t_3-t_2)P_{A_2}U(t_2-t_1)P_{A_1}\eeq
Where $U(t)$ is the unitary evolution operator that relates to the Hamiltonian via $U(t)=\exp (-iHt)$, and $P_{A_i}$ is the subspace that history $A$ lied at time $t_i$. This class operator corresponds to the history that the system is at $P_{A_1}$ at $t_1$ \emph{and} at $P_{A_2}$ at $t_2$ etc. The projection operators can be at any subspace of the Hilbert space. The quantum measure is then defined to be

\beq\mu(A)=\Tr(C^\dagger_A\rho C_A)\eeq
For cases where there is a well defined time, such as in non-relativistic quantum mechanics, the two definitions are equivalent and one uses the more convenient one. For finite moments of time histories, the operator expression is usually easier to deal with\footnote{Introducing more moments of time is simply a fine graining of the previous histories.}.

Note, that the quantum measure is closely related with the decoherence functional \cite{Decoherent Histories}, since it arises from the diagonal parts of the latter.

Since the quantum measure is a non-negative function that is also normalised ($\mu(\Omega)=1$), one could be tempted to interpret the quantum measure as probability. However this is not possible, due to interference. The additivity condition of probabilities\footnote{Additivity of disjoint regions of the sample space.}, is not satisfied:

\beq \mu(A\sqcup B)\neq\mu(A)+\mu(B)\eeq
However a weaker condition holds that shows that there is no three-paths interference, that cannot be deduced from pairwise interference:

\beq\label{three-interference} \mu(A\sqcup B\sqcup C)=\mu(A\sqcup B)+\mu(A\sqcup C)+\mu(B\sqcup C)-\mu(A)-\mu(B)-\mu(C)\eeq
In other worlds, the quantum measure is fully determined once we know the quantum measure of single histories and of pairs of histories. The specific expression is given for example in \cite{surya2010}:
\beq\mu(A=\{h_1,h_2,\cdots,h_n\})=(2-n)\sum_{i=1}^n\mu(h_i)+\frac12\sum_{i,j=1}^n\mu(\{h_i,h_j\})\eeq
Recently, experiments \cite{three slits} have confirmed that indeed the condition of eq. (\ref{three-interference}) holds in nature.

To fully characterise a histories theory, we need the following triplet: The histories space, the events algebra and the quantum measure $(\Omega,\mathcal U,\mu)$.

A \emph{partition} $\mathcal P$ of the histories space $\Omega$, is a collection of events $\{P_1,P_2,\cdots\}$ such that $P_i\cap P_j=\emptyset$ for all $i\neq j$ and $\bigcup_i P_i=\Omega$. It is an exhaustive and exclusive collection. Each event of the partition is called \emph{cell}. A \emph{coarse-graining} of a histories theory $(\mathcal P, \mathcal U_{\mathcal P},\mu_{\mathcal P})$ is defined in the following way:

 \begin{enumerate}
 \item The cells of a partition $\mathcal P$ are the elements of the coarse-grained histories space.
 \item The Boolean algebra generated by addition and multiplication of these cells $\mathcal U_{\mathcal P}$, is the coarse grained event algebra.
 \item The quantum measure defined on them should be compatible with the fine grained quantum measure, i.e. $\mu_{\mathcal P}(A)=\mu(A)\textrm{ for all }A\in\mathcal U_{\mathcal P}$
 \end{enumerate}

\subsection{The trouble with quantum theory}\label{2.4}

It is already clear, that since the quantum measure cannot be understood as probability, the picture of classical stochastic physics presented in the previous section might need modification. However, the main reason why one cannot interpret the quantum measure in the same way as the classical measure, appears from consideration of the precluded events. Note that from now on, when we mention precluded events we mean that they have \emph{quantum} measure zero.

In many cases (see below for examples), one can find a collection of precluded events $\mathcal Z=\{P_1,P_2,\cdots\}$, where $\mu(P_i)=0$, that also cover the full histories space $\Omega$, i.e. that $\bigcup_i P_i=\Omega$. We will call such collection of precluded events as a \emph{zero cover}. Note, that since in general $P_i\cap P_j\neq\emptyset$, this collection does \emph{not} constitute a partition.

However, as we mentioned earlier, in order to be able to construct deductive arguments, the truth value of \emph{any} subset of a precluded set has to be ``False''. Having the full $\Omega$ covered with precluded sets, imply that no single history $h_i\in\Omega$ can have a truth value ``True''. This also leads to no homomorphic map $\phi\in\mathcal M$ existing.

\begin{figure}[h]
\center\scalebox{0.30}{\includegraphics{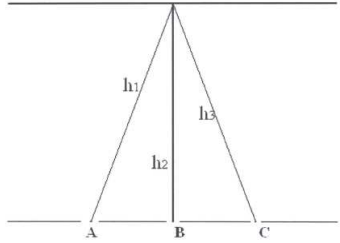}}\caption{Three
slits, $h_1$ is the history that passes from slit $A$ and hits the
central point on the screen, etc}\label{3 slits}
\end{figure}

We will first consider a simplified example, with a three-slits interference experiment. Consider a point at the screen, where crossing slit A destructively interferes with crossing through slit B and slit B destructively
interferes with slit C, i.e. $\mu (\{h_1,h_2\})=0, \mu(\{h_2,h_3\})=0$, however $\mu(\{h_1,h_2,h_3\})\neq0$ (see Figure \ref{3 slits}).

Since the quantum measure for hitting that point on the screen is non-zero, we know that particles do hit the screen at this point some times. In these cases we run into paradox, by assuming that the particle followed a single trajectory. Crossing from any slit, is a subset of an event of quantum measure zero and thus cannot occur no matter what possible hidden variables determine the realised trajectory. However, in this example, one can say that we did not cover all $\Omega$ with precluded events, because the particle could have been detected at other points of the screen where the above analysis does not hold. Even though the nature of the problem that arose in this example was essentially the same as that for zero covers, this example is not a proper zero cover.

We now give, an example of a proper zero cover for a single qubit with initial state $|0\rangle$. We measure it in three-moments-of-time\footnote{Having more moments of time, is simply a fine-graining and does not alter the coarse-grained cover we give.}. We assume that the qubit evolves in the following (unitary) way:

\beq U=\frac1{\sqrt2}\left(\begin{array}{c c}1&i\\ i&1\end{array}\right)\eeq which can arise for some particular Hamiltonian $H$ and time interval $t$ between measurements.


We evolve and measure in the $\{|0\rangle,|1\rangle\}$ basis, three times. The amplitudes for different histories are given by the expression using the class operators $|\psi_{final}\rangle=C_\a|0\rangle=P_{t_3}UP_{t_2}UP_{t_1}U|0\rangle$, with $P_{t_i}$ being the projection either $|0\rangle\langle0|$ or $|1\rangle\langle1|$ depending on what was measured at the moment $t_i$:

\bea
\a(h_1=000)=\frac1{2\sqrt2} &,& \a(h_2=001)=-\frac1{2\sqrt2}\nonumber\\
\a(h_3=010)=-\frac1{2\sqrt2} &,& \a(h_4=011)=-\frac1{2\sqrt2}\nonumber\\
\a(h_5=100)=\frac1{2\sqrt2}i &,& \a(h_6=101)=-\frac1{2\sqrt2}i\nonumber\\
\a(h_7=110)=\frac1{2\sqrt2}i &,& \a(h_8=111)=\frac1{2\sqrt2}i
\eea
The labels indicate outcomes of the $\{|0\rangle, |1\rangle\}$ measurements from right to left. For example $h_4$ is the history that we get the outcome $|1\rangle$ in the two first measurements and $|0\rangle$ in the third.
This leads to the following zero cover (collection of precluded events that their union is $\Omega$):

 \beq \mathcal Z=\{\{h_1,h_2\},\{h_1,h_3\},\{h_1,h_4\},\{h_5,h_6\},\{h_6,h_7\},\{h_6,h_8\}\}\eeq
This example is simple and very general, since with little modifications, it can hold for any Hamiltonian for the qubit (choosing different moments of time).

More complicated examples can also be constructed for all systems (initial states, Hamiltonian). The Kochen-Specker theorem \cite{KS thm}, when casted in the histories language, is precisely a special case of a zero cover (see \cite{surya2010,Dowker2008}). It is the existence of these zero covers, that rules out the possibility of the reality being a single-history and thus rules out hidden variables theories. We have to stress however, that the conclusion holds, provided one accepts as starting point the Feynman path integral as the amplitude for all histories, being fine or coarse grained\footnote{In De Broglie-Bohm theory, the problem of zero covers is avoided, by deviating from the Feynman amplitudes in general and only agreeing when necessary to recover standard non-relativistic QM results.}.

One attempt to overcome the zero covers paradox, would be to avoid taking \emph{all} events with quantum measure zero as precluded, i.e. allow for some events attaining truth value ``True'', even if they have quantum measure zero. This runs to, at least two problems. One is that we need a physical/natural criterion for when an event of quantum measure zero is actually precluded and when it is not. The second comes from consistent histories approach, and the observation, that quantum measure zero events, decohere (and thus behave classically) with their complement. Having, a zero cover, leads necessarily either to a context-dependent view or to the need of selecting a preferred classical domain for the consistent histories (see section \ref{4.1}).

Other attempts involve modifying one of the structures we had in classical physics ($\mathcal U, \mathcal T, \mathcal M$).

\begin{enumerate}
\item Instead of considering the full set of possible questions the event algebra $ \mathcal
U$ (subsets of $ \Omega$), to somehow select a preferred set of
classical questions, in other words a subalgebra of the event
algebra. This is essentially done in the consistent histories approach \cite{Decoherent
Histories}. However, this approach allows an
arbitrary choice of which ``Boolean subalgebra'' to consider (see
the critic in \cite{DoKe}). In this sense, fails to give a
satisfactory account of what actually occurs, unless some physical
principle is discovered that selects a preferred classical domain (see section \ref{4.1}).

\item Alter the space of truth values (and the algebra associated) $\mathcal T$. Instead of using a two values Boolean logic ($\{\textrm{True, False}\}$)
one could use truth valuation maps that take truth values on a
subobject classifier of a category, and the associated algebra is a
Heyting algebra (see for example \cite{isham-topos}). The resulting logic is intuitionistic logic
which is deductive logic\footnote{Proofs by contradiction are not allowed, and only
constructive/deductive proofs are acceptable.} but most importantly,
contextual (the truth value, depends on the context/question asked). One can also attempt to interpret consistent histories in this contextual view \cite{isham-topos-histories}.

\item Finally, one could alter the set of allowed maps $\phi$ from $ \mathcal U$ to $ \mathcal T$, while keeping both the event algebra and the truth values algebra, same as in classical physics.
This is achieved by weakening the requirement that $\phi$ is a homomorphism. This is essentially done in the coevent formulation which is the topic of this paper.
\end{enumerate}

\section{The coevent formulation}\label{3.0}

We are now in position to give the technical definition of the coevents formulation. The approach was pioneered by Sorkin \cite{coevent1,coevent2} and one can find older reviews in \cite{Sorkin2010,Ga}. There are two, essentially equivalent, ways of viewing the coevents formulation. Here we will introduce both, starting with what appears as more intuitive, leaving as second the one that was historically developed first. In particular, the first view, is that the reality (ontology of the theory) is a coarse-grained history, in direct analogy with the picture in classical physics, that reality was a fine-grained history $h_i$. The second view, is the dual one, that considers the reality as being the (generalised) valuation map $\phi$, which is in general anhomomorphic, and it is analogous with the view in classical physics, that reality is a homomorphic map $\phi_{h_i}$ that is the characteristic function of the history $h_i$.

\subsection{The coarse-grained history view}\label{3.1}

Instead of viewing as possible realities the fine-grained histories $h$, elements of $\Omega$, we could allow as potential realities, coarse-grained histories, i.e. (as starting point) any event $R$ element of $\mathcal U$. However, it is clear from classical physics, that this would lead to allowing potential realities that are unphysical. We need to impose conditions, that among other things, will reproduce the classical picture for the case of classical measure. There are two such requirements.

\begin{enumerate}

\item The allowed potential realities (events), are not precluded, and moreover, they are not included in any precluded event. We define an event $A$ to be \emph{non-preclusive} if

    \beq\label{non-preclusive}\nexists\quad P\textrm{ such that }A\subseteq P\textrm{ and } \mu(P)=0\eeq
    The set of all non-preclusive events we call it $\mathcal {NP}$. We require that an allowed reality $R$ is a non-preclusive event ($R\in\mathcal{NP}$).

    This condition is satisfied by potential realties in classical physics. However, for a classical measure, all subsets of a measure zero set, are also measure zero sets. It follows, that in classical physics is sufficient to require that potential realities $h_i$ (single-histories events) are not precluded ($\mu_c(h_i)\neq 0$).

    For a quantum measure, it is possible to have an event $A$ that is not precluded ($\mu(A)\neq 0$) but \emph{is} subset of a precluded event $P$. As we have seen earlier, we wish to rule out such events from potential realties, and this is done by requiring the allowed coevents to be non-preclusive (rather than being simply not precluded). To this point, we stress again the reason to rule out such events. We wish to recover deductive logic, and we cannot have an event given truth value ``True'', while one of its supersets is precluded and thus given the truth value ``False''.

\item The second condition is that the potential realities $R$ are the finest grained events that are non-preclusive, i.e. obey eq. (\ref{non-preclusive}). Mathematically:

    \beq\label{max-detail} R\in\mathcal{NP}\textrm{ and }\nexists\quad A\in\mathcal{NP}\textrm{ such that }A\subset R\eeq
This requirement states, that from all the non-preclusive events, we consider as potential realities \emph{only} those that give the most detailed description. This is called the ``maximum detail'' condition. It is precisely this condition, that in the case of a classical measure, restricts the possible realties to be fine-grained histories that are not precluded, and thus agree with the picture developed in the previous section. For a quantum measure, the condition of eq. (\ref{non-preclusive}) forces us to consider potential realities that are not single-histories, as we will see in examples below.
\end{enumerate}
The potential realities are events that are non-preclusive and of maximum detail and we will call them \emph{covents}\footnote{Note that the term ``coevent'', is used usually for the dual picture and refers to the corresponding maps, while the term ``support of the coevent'' is used for the events described here. We will not adopt this convention in this paper, but will comment on this later.}. The set of all coevents, given a quantum measure, is $\mathcal R$.

In order to find the set of possible coevents, we only need to know which events have zero quantum measure. Let us see, which are the allowed coevents for the examples of zero covers we gave in the previous section.

In the three-slit example, the precluded events, were $\{h_1,h_2\}$ and $\{h_2,h_3\}$. The non-preclusive events (those not included in precluded events) are $\{h_1,h_3\}$ and $\{h_1,h_2,h_3\}$ which are the elements of $\mathcal {NP}$. From those, only the event $\{h_1,h_3\}$ satisfies eq.(\ref{max-detail}) since the other event $\{h_1,h_2,h_3\}$ has as subset the non-preclusive event $\{h_1,h_3\}$. We therefore have a single coevent as potential reality and $\mathcal R=\{\{h_1,h_3\}\}$.

In the qubit example with the three-moments-of-time, we have many precluded events, namely:

\bea & &\{h_1,h_2\},\{h_1,h_3\},\{h_1,h_4\},\{h_5,h_6\},\{h_6,h_7\},\{h_6,h_8\},\{h_1,h_2,h_5,h_6\},\nonumber\\
& &\{h_1,h_2,h_6,h_7\},\{h_1,h_2,h_6,h_8\},\{h_1,h_3,h_5,h_6\},\{h_1,h_3,h_6,h_7\},\nonumber\\& &\{h_1,h_3,h_6,h_8\},
\{h_1,h_4,h_5,h_6\},\{h_1,h_4,h_6,h_7\},\{h_1,h_4,h_6,h_8\}
\eea
One can find that there are the following six coevents:

\beq
\mathcal R=\{\{h_2,h_3\},\{h_2,h_4\},\{h_3,h_4\},\{h_5,h_7\},\{h_5,h_8\},\{h_7,h_8\}\}
\eeq
This is an example where the potential realities, are not single-histories and moreover there are many (six) such potential realties.

After having defined the new ontology of the theory, as being that of a coevent, we wish to be able to make assertions. Given one realised coevent, which physical questions/events occur, i.e. take truth value ``True''? The truth value ``True'' is given to all  events that have the realised coevent $R$ as a subset and the rest take truth value ``False''. However, this might lead us to a truth valuation map $\phi$ that is \emph{not} homomorphic. We will analyse this in the second view of the coevent formulation that follows.

Note that the above, analysis holds for finite histories space $\Omega$. Maintaining this view for infinite histories space is possible but needs more further analysis in the definitions in order to be rigorously formulated.

\subsection{The anhomomorphic valuation view}\label{3.2}

In classical physics, we mentioned a dual view of reality. Instead of considering the single-history $h$ as the ontology of the theory, one could consider the homomorphic truth valuation map $\phi_h$ as the reality. These valuation maps are characteristic function of single histories $h$, i.e. gives truth value ``True'' to any event $A$ that includes the history $h$, and ``False'' to all the other events.

\beq\phi_h(A)=1\textrm{ iff }h\in A\eeq
The second view of the coevents formulation, is to try and generalise the truth valuations $\phi$'s. It should be done in such a way, that is compatible with the paradoxes arising from using a quantum rather than a classical measure. In particular it should be compatible with the presence of zero covers. The modification that we adopt, is weakening the requirement that the truth valuation maps are homomorphisms between the Boolean algebras of $\mathcal U$ and $\mathcal T$. In particular we still require the maps $\phi$ to be multiplicative i.e. obey eq.(\ref{multiplicativity}), but no longer require them to be additive.

\bea
\phi(A\cdot B)&=&\phi(A\cap B)=\phi(A)\phi(B)\nonumber\\
\phi(A+B)&\neq&\phi(A)+\phi(B)\eea
 We should note, that the first attempt of weakening the homomorphism to allow for the zero covers, was keeping the additive property and dropping the multiplication \cite{coevent2} but it was ruled out by some gedanken experiments that provided counter examples where no such map existed at all. There are also other attempts to weaken the homomorphism. All different attempts were coined ``schemes'' and the one we have adopted and present in this paper, is the multiplicative scheme. Other than concrete examples that alternative schemes fail, there are several advantages in adopting the multiplicative scheme: (i) being able to have the coarse-grained view, (ii) being guaranteed that there always exists allowed maps no matter which is the quantum measure, (iii) recovering deductive logic and (iv) having a unique classical domain (we will analyse (ii),(iii) and (iv) later).

It can be shown, that all multiplicative maps, are in fact characteristic maps of some (non-trivial) event $S$

\beq\phi_S(A)=1\textrm{ iff} S\subseteq A\eeq
We call this event $S$ \emph{support} of the multiplicative map. Two things to note. First, is that the choice of allowed maps as characteristic maps is very similar with what was done in classical physics with the only difference that we now allow for characteristic maps of events that may consist of many fine grained histories $A=\{h_1,h_2,\cdots\}$. Second thing to note, is that precisely this property of multiplicative maps allows us to have both views of the coevent formulation. There is a one-to-one correspondence of anhomomorphic (but multiplicative) maps and coarse-grained histories (their supports).

Other than requiring the allowed coevents to be multiplicative, we need to impose two further conditions to potential realties, as it was done in the coarse-grained history view.

\begin{enumerate}

\item The maps are preclusive, in other words they give truth value ``False'', for any event that has zero quantum measure

     \beq \mu(P)=0\Rightarrow \phi(P)=0\eeq
     This condition together with the multiplicativity condition, guarantees that an event $A$ being subset of a precluded event $P$, gets truth value ``False''. This plays the same role as the first requirement in the previous section (non-preclusive).

\item The condition that corresponds to the maximum detail condition of section \ref{3.1}, is that the map needs to be primitive. We say that a multiplicative map $\psi$ dominates a multiplicative map $\phi$ if

\beq \phi(A)=1\Rightarrow\psi(A)=1\quad\forall\quad A\in\mathcal{U}\eeq
A preclusive multiplicative map is \emph{primitive} if it is not
dominated by any other preclusive multiplicative map. For a classical measure, the primitivity condition, leads to maps that are homomorphisms and thus correspond to characteristic maps of single histories, as expected.

\end{enumerate}

The potential realities are (also) called coevents, and are valuation maps that are (i) multiplicative, (ii) preclusive and (iii) primitive.

We should make here a comment on the terminology used to avoid confusion. As coevent, we defined both the multiplicative, preclusive and primitive map $\phi_R$ and the support of this map $R$ which is a non-preclusive, maximum detail event. In other words we called coevents, the potential realities of the coevents formulation, no matter if we adopted the coarse grained view or the anhomomorphic valuation view. In literature, the term coevent was used only for the (in general anhomomorphic) maps. However, when restricting attention in the multiplicative scheme as we are in this paper, the coarse grained history view is equivalent. In this case, even referring to the anhomomorphic  map is easier  done using its support.

Since, we no longer have homomorphism between the events Boolean algebra $\mathcal U$  and the truth values Boolean algebra $\mathcal T$ the laws of inference do not follow directly. The main deductive inference rule, is the modus ponens that essentially means if $A$ is ``True'' and $A$ implies $B$, ($A\rightarrow B$), then $B$ also is ``True''. In \cite{MP} it was shown, that modus ponens, is applicable for multiplicative maps. In particular, it was shown that if we require that our valuation maps obey modus ponens, we are necessarily led to multiplicative maps. This is one further motivation for the choice of the multiplicative scheme. It is clear however, that due to the fact that we no longer require the maps to be homomorphisms, some of the structure of the Boolean algebra is lost. In particular, we cannot use proofs by contradiction. There exist events $A$ such that both $\phi(A)=0$ and $\phi(\neg A)=0$.

 In reality, one of the possible coevents is realised. Given that coevent, we should be able to answer whether any given question/event is realised (``True'') or not realised (``False''). As long as the event in question $A$, either includes the realised coevent $R_1$, i.e. $R_1\subseteq A$ or does not intersect it at all $R_1\cap A=\emptyset$, it is easy to see that either $\phi_{R_1}(A)=1$ or $\phi_{R_1}(\neg A)=1$. However, if the event $A$ intersects non-trivially $R_1$, then both $A$ and $\neg A$ take truth value ``False''.

 For example, in the three-slit case (see Figure \ref{3 slits}), we have a single allowed coevent $\{h_1,h_3\}$ which is realised. Asking the question ``did the system cross from any of slit A or B'' ($A=\{h_1,h_2\}$) gets answer ``No''. However, answer ``No'' gets the question ``did it cross slit C'' ($\neg A=\{h_3\}$). This type of paradox is natural for the quantum world, but should not persist in classical world (see later). In other words, our anhomomorphic logic, allows for deductive proofs but not for proofs by contradiction which is precisely what happens in intuitionistic logic\footnote{Note however, that our logic is \emph{not} intuitionistic.}. The latter is partly the motivation for the field of ``constructive mathematics'' where mathematics are reformulated using solely deductive (constructive) proofs.

Note here that while the rule
\beq\label{contradiction}\phi(A)=0 \Rightarrow \phi(\neg A)=1\eeq
 does \emph{not} hold in the coevents formulation, and we cannot use proofs by contradiction, the opposite rule holds. Namely, if $A$ is ``True'' its complement is indeed ``False''.
 \beq\label{no name}\phi(A)=1\Rightarrow \phi(\neg A)=0\eeq

\subsection{Remarks}\label{3.3}

The problem in applying the classical picture of histories in the quantum world arises because we have a quantum rather than a classical measure. The quantum measure allows the existence of zero covers. We therefore can no longer maintain the picture of reality as a single fine-grained history or equivalently as a homomorphic map.

We are led to extend the potential realities from single fine-grained histories to coevents (either viewed as a coarse-grained histories or as multiplicative maps). After this generalisation, we are guaranteed to have some potential realities that are compatible with \emph{any} quantum measure (see \cite{surya2010}).

The initial conditions and dynamics fix the quantum measure. From that we obtain the precluded events which determine the allowed coevents. Given the quantum measure, we have, in general, many allowed coevents. In this sense, quantum theory appears as a generalisation of classical stochastic physics rather than generalisation of classical deterministic physics. Measurements are treated in a very similar way as in stochastic classical physics. They are used to rule out some of the potential realities/coevents, alas without altering the system (unlike standard Copenhagen quantum mechanics). The new ontology of quantum theory, is that of a coevent, is not affected by measurements (in the way it does in Copenhagen quantum mechanics) and is not contextual. We can therefore conclude that the coevents formulation is realist.

A final point to raise here, concerns the role of logic. As we have seen, we retain deductive proofs. However, the precise form of logic, i.e. when other rules of inference can be applied, depends crucially on the set of possible coevents, which in their turn depend on the particular quantum measure. It therefore appears that the logic of quantum world is no longer fixed, but becomes dynamical. To quote Sorkin ``Logic is to quantum, as geometry is to gravity'' \cite{Sorkin2010}.

\section{Developments}\label{4.0}

Having given the definition of the coevent formulation, we will now proceed analysing certain important aspects.

\subsection{Classical domain and consistent histories}\label{4.1}

A very important question for quantum theory of a closed system is how one recovers, at some level/coarse-graining, classical physics. The finest grained description is not expected to behave classically, and thus some counter-intuitive properties such as the anhomomorphisms described earlier (e.g. end of section 3.2) are natural. We expect, though, that if we coarse-grain sufficiently, the arising structure behaves classically. In our case, the non-classical behaviour is the anhomomorphisms of the valuation maps. We call classical coarse graining one that all the allowed coevents, elements of $\mathcal R$ give rise to homomorphism between the coarse-grained event algebra and the truth values Boolean algebra.

Given a quantum measure $\mu$ we have many ``allowed'' coevents $R_i\in\mathcal R$. A classical coarse
graining $\mathcal C$ is a coarse-graining (see end of section 2.3) such that \emph{all} the coevents $R_i$ give rise to homomorphism. To do so, all coevents $R_i$ should be subsets of some cell of the partition:

\bea \mathcal{C}&=&\{C_1,C_2,\cdots\}\textrm{ partition}\nonumber\\
\textrm{such that} &\forall& R_i\in\mathcal R\quad\exists\quad j \textrm{ such that }R_i\subseteq C_j
\eea

One of the most important results, is that there exists a \emph{unique} finest grained classical
partition, which we will call \emph{principle classical partition}. This means that \emph{all} classical partitions arise as coarse grainings of this unique finest grained partition (see \cite{YGT-PW1} the Appendix). The construction of the principle classical partition is done as follows\footnote{A more rigorous proof can be found in the original reference.}:

\begin{enumerate}

\item If no pair of coevents intersects $R_i\cap R_j=\emptyset$,
the finest partition, is one that has each of $R_i$ as
a cell $C_i$ of the partition, and the rest of the history space $\Omega\setminus\bigcup_iR_i$ is
covered by single-history cells $C_k=\{h_k\}$ for all $h_k\in\Omega\setminus\bigcup_iR_i$.

\item If the coevents overlap
($R_i\bigcap R_j\neq\emptyset$), we take the union of the intersecting coevents and consider this as a cell
$C_i$ of the classical partition\footnote{We need to take the transitive closure under intersection. If for example $R_1$ intersects non-trivially with $R_2$ and $R_2$ with $R_3$, we need to consider the union of all three coevents, even if $R_1\cap R_3=\emptyset$.}. The rest space $\Omega\setminus\bigcup_iR_i$ is again
covered by single-history cells $C_k=\{h_k\}$ for all $h_k\in\Omega\setminus\bigcup_iR_i$.

\end{enumerate}
Any measurement we can carry out with ``classical'' apparatuses, will correspond to some coarse-graining of the principle classical partition. Therefore, it is not possible to distinguish between two intersecting coevents with any classical measurement. One could view these ``fat'' coevents (union of the intersecting ones), as more physical, or treat this property as a short of ``coevents uncertainty''.

The importance of this result, for histories formulations, becomes more apparent if we compare the situation with the consistent histories approach \cite{Decoherent Histories}. The consistent histories approach, with the use of the decoherence functional, finds partitions $\mathcal P_i=\{P_1^i,P_2^i,\cdots\}$ of the histories space $\Omega$, such that there is no interference between pairs of cells of the partition. The induced quantum measure becomes classical\footnote{The quantum measure defined on the coarse graining event algebra, is additive and thus classical.}. Each of these partitions, is called a \emph{consistent set}, and one could attempt to view each consistent set as a classical domain. However, there does \emph{not} exist a single finest grained consistent set that all the others arise from. This leads to properties in one consistent set being incompatible with properties on another, and we are forced to one of the following two solutions. Either with some physical mechanism (not known yet) to select a preferred consistent set, or to adopt a contextual view, that the answers to propositions depend on which consistent set one wishes to refer to (for further details see \cite{DoKe}).

We have, so far, examined the meaning, existence and uniqueness of a classical domain. What remains, is to examine particular physical examples and how this classical domain emerges from the dynamics and whether it agrees with our classical intuition (pointers in a lab being classical, etc). This is essentially matter of calculation (see section 7 of \cite{Sorkin2010}) and is subject of ongoing research.

\subsection{Probabilities}\label{4.2}

The analysis we did so far, concerned the potential realities. However, there is much more to a physical theory, than solely stating which are the possible realities. One needs to be able to make predictions, and in quantum theory, these predictions are in general probabilistic. We should be able, for example, to recover the double slit pattern, if we consider as system a large number of particles crossing the double slit apparatus.

However, the concept of probabilistic prediction for a single non-repeatable closed system is problematic already in the classical level. For example a statement such as ``the universe has property $A$ with probability 1/3'' cannot be falsified by any experiment. Both results, possessing or not possessing the property, are compatible with the ``prediction''. This fails the criterion for a scientific proposition, which is that there should exist possible outcomes that falsies the assumption (see Popper). In our case, the system is closed and ideally non-repeatable (we construct such a theory to be able to treat the whole universe as a quantum system). We therefore need to understand the meaning of probability and its connection to the physical world, and to do so we have to resort to the founders of probability theory.

We can see, that in a non-repeatable system, the only testable predictions, are those that concern things occurring with probability one or zero. We have also seen in section \ref{2.2} that in the limit of infinite identical copies, one can recover probabilities from precluded events\footnote{Note that the expression at that section, concerns a classical probability measure, and not a quantum measure. However, recovering probabilistic predictions from approximately precluded events can be done for a quantum measure as well, as was done for example, for the double slit example in \cite{YGT-PW2}.}. Considering the above, one is lead to the Cournot principle \cite{Cournot} (see also Kolmogorov's view  \cite{Kolmogorov} and the references in \cite{YGT-PW1}):

\begin{quote}
In  a repeated trial, an event $A$ singled out in advance, of small measure ($\leq\epsilon$), rarely occurs.
\end{quote}
There exist a stronger formulation of the principle where the ``rarely occurs'' is replaced with ``does not occur'', but we adopt the one above. First thing to note in this definition, is the term ``small measure''. In the coevents formulation we will call such events \emph{approximately precluded}. It is clear, that this is a subjective choice, and depending on how small $\epsilon$ is, different prediction may arise. Once the theoretician makes the choice of $\epsilon$, the rest follows. Second thing to stress, is the importance of selecting the event $A$ in advance. Before exploring the latter, we will give an example of how Cournot's principle does work.

Consider tossing a coin 1000 times, and start with the assumption that the coin is fair. Any sequence of 1000 outcomes heads/tails is a fine grained history $h_i$. Suppose we select the event $A$, which is the history that has all outcomes to be heads\footnote{The argument holds for other non-trivial events, such as the event that has all series of outcomes, with more than 990 heads outcomes.}. The probability (with the assumption of a fair coin) is clearly negligible ($1/2^{1000}$). We can thus predict that the event $A$ will not occur, for most reasonable choices of $\epsilon$. However, it is conceivable that such an event does occur. In that very rare occasion, that such event occurs, we would be led to believe that the coin was not fair and thus falsify our initial assummption\footnote{Actually, we might even check whether the coin had at both sides head!}. This is precisely what is done in actual experimental practise, with the only exception, that if our belief for the assumption is strong, we may convince ourselves to further repeat the experiment if that is possible. In a fundamentally closed system however, such option is not available.

We return now to the importance of pre-selecting the event to be tested. All the time in nature, things of arbitrary small probability do occur. In the earlier example, of tossing a coin 1000 times, any sequence of outcomes heads/tails, has extremely small measure and it is identically the same with the outcome ``1000 times heads'' that we dismissed earlier. However, the difference between the two is the following. In the one case (1000 times heads) we could ask whether this will occur before carrying the experiment. For some other random sequence of outcomes (the one that actually happened), we did not ask in advance. If we had given a sequence of 1000 outcomes in advance, and then this precise sequence did occur when we carried out the experiment, we would be very sceptical and troubled, and likely assume that somehow there was a correlation of our mind and the tossing of the coin. If the number of tosses were greater than 1000, then we would be even more surprised. In general, we would dismiss the assumption of a fully fair coin and attempt to find which was the pattern that gave this series of outcomes (in that example, it would be easy to perform few more measurements to test the hypothesis, but this is not always an option for all physical systems).

We therefore see that in the use of the weak Cournot principle, there is a dichotomy between the ontology and the prediction of the theory. There exist things of small measure, that despite being conceivable realities, one predicts that they will not occur. And if they do occur, the theoretician is led to the wrong conclusion that his assumptions were wrong\footnote{For example, it is conceivable that the fact that the universe in large scale looks homogeneous is \emph{not} due to a common cause, but it happened randomly. In our construction of a theory of cosmology, we dismiss such coincidence and try to justify it with the existence of a common cause.}.

Note here, that the Cournot principle was supported by many of the founders of probability theory, such as Bernoulli, Cournot, Markov, Borel, Levy, Kolmogorov and many others. Borel stated that Cournot's principle ``is the \emph{only} law of chance'', while Levy stated that ``Cournot's principle is the \emph{only} connection between probability and the empirical world''. The use of the principle in recent times has fallen out of fashion, mainly due to the dominance of the subjective view of probabilities such as the approach of De Finetti where probabilities are treated as betting strategies. However, both the objective view of probabilities and the use of precluded sets as the essence of prediction, suits very well the coevents formulation. This is because the coevents formulation deals with a single closed system, adopts a realist view and uses the preclusion as the central feature in selecting the potential realities/coevents.

In coevents formulation the dichotomy between ontology and predictions persists.  In particular, the ontology of the theory is the coevents themselves, while to recover probabilistic predictions one takes the quantum measure and applies the weak Cournot principle as described above. In \cite{YGT-PW2} for example, one can see how a discrete version of the double slit pattern is recovered using the quantum measure and applying the Cournot principle\footnote{Note, that using any other classical measure, would \emph{not} recover the double slit pattern.}.

\subsection{The role and status of the initial state}\label{4.3}

In the coevent formulation, the role of the initial state, is that it affects the quantum measure and thus determines both the allowed coevents, and the probabilistic predictions (see previous section). An interesting question to ask, would be whether the initial state characterises the system (universe) in the following way. Given the coevent realised\footnote{Assuming for the moment that such thing is possible.}, can we uniquely determine which was the initial state? Or in other words, is a universe with initial state $\Psi_1$ different from one with $\Psi_2$? The answer to those questions is closely related with the interpretation of the (initial) state, as was analysed by Pusey Barrett and Rudolph in \cite{PBR}. If one views different states as ontological different entities, the answer is ``yes'', while if one views them as statistical distributions, the answer is ``no''.

Given an initial state $\Psi_1$, we can find the set of allowed coevents $\mathcal R_1=\{R_1^1,R_2^1,\cdots\}$ and similarly for initial state $\Psi_2$, we have $\mathcal R_2=\{R_1^2,R_2^2,\cdots\}$. If there are no common covents $\mathcal R_1\cap\mathcal R_2=\emptyset$ for any pair of initial states $\Psi_1,\Psi_2$, then we can say the universe starting with $\Psi_1$ is different from that with $\Psi_2$. Otherwise, it is possible that the actual realised coevent is compatible with two (or more) initial states.

For the coevents formulation, in \cite{Wa2012} was given evidence\footnote{See the appendix of \cite{Wa2012}. Note however, that the particular gendanken experiment proposed in \cite{PBR} did not fully carry over to the coevent formulation, due the failure to satisfy in histories approaches one of their assumption.} that if one looks at sufficiently fine-grained description, the sets of coevents starting from different pure states is disjoint, i.e. the universe of $\Psi_1$ \emph{is} different from that of $\Psi_2$. It follows that one is at least in principle, able to retrodict uniquely the initial state. While the initial state per se, has no ontological status in histories formulations, it obtains such status indirectly. Any possible reality/coevent, can arise from a unique initial state, and in this sense the initial state becomes a property possessed by the system. This observation has also interesting consequences for the field of (quantum) cosmology.

\subsection{Other developments and outlook}\label{4.4}

There are several other questions that one can ask in relation with the quantum measure and its interpretation in the coevents formulation. Here we will briefly mention some works that have been done giving references.

One interesting thing to analyse, that is closely related with the question asked in the previous section, is which properties of our system can we deduce from the coevents themselves, or else which are the properties possessed by the coevents. One could wonder for example, what is the meaning of energy, for a given coevent, i.e. if our ontology is just a subset of histories. In a toy model of a three-sites hopper Sorkin showed that the system ``does know its own energy'' from simply looking the coevents \cite{Sorkin_Athens}, and we can treat the energy of the site-hopper as a property the coevents have.

Other works that have been done, is the attempt to formally extend the quantum measure for infinite histories space. The problems that arise can be found in \cite{DoJoSu} and directions on how to overcome them in \cite{Sorkin2011}. The derivation of a Hilbert space structure starting purely from histories is given in \cite{coevents_Hilbert} and the concept of quantum integration and its use is developed in (for example) \cite{Gudder}.

There are several direction to further explore. One direction is exploring the conceptual issues that could arise from (i) considering composite systems, (ii) time extending the coevents (in other words, the effects that fine-graining the quantum measure has, to the set of allowed coevents) and (iii) considering large systems that classicality should emerge. Technical issues to be explored, are: (1) the extension of quantum measure and rigorously founding the formulation for infinite histories space and (2) finding ways to compute the allowed coevents for complicated systems (for the moment, it is very hard to do so, even for histories space with only few histories).

\section{Summary and conclusions}\label{5.0}

We presented the coevents formulation of quantum theory. It is a realistic, histories formulation that applies to closed quantum systems and is well suited for quantum gravity. The three-fold structure of applying histories was presented, where we have (i) the space of histories $\Omega$ along with the event algebra $\mathcal U$, (ii) the space of truth values $\mathcal T$ and (iii) the space of allowed truth valuation maps $\mathcal M$. The latter in classical physics is the homomorphic maps $\phi$ from $\mathcal U\rightarrow\mathcal T$. The role of the (classical) measure, the difference of classical deterministic and stochastic physics and the importance of precluded events were analysed. The replacement of the classical measure with quantum measure led us to the need to alter the picture. This was due to the existence of zero covers which ruled out the single-histories as potential realities.

As a solution we introduced the coevents formulation, which can be viewed in two different ways. The new ontology, is either coarse-grained histories (non-trivial events) $R_i$ that are non-preclusive and are of maximum detail or multiplicative maps $\phi_{R_i}$ (in general anhomomorphic), that are preclusive and primitive. The two pictures are equivalent, since there is a one-to-one correspondence between events and multiplicative maps, and the further conditions are also equivalent. With the coevents as ontology, we avoid the paradoxes described, that arose from the existence of zero covers. This is the case, because there is an existence theorem showing that given a quantum measure, there always exist non-trivial coevents (see \cite{surya2010}). The resulting logic, while it is not Boolean, is deductive (the modus ponens inference rule holds \cite{MP}). Proofs by contradiction are not, in general, possible due to the failure of eq. (\ref{contradiction}). However, the opposite rule given by eq. (\ref{no name}) holds.

The coevents formulation, appears as a generalisation of classical stochastic physics, where there are many potential realities. The ontology (coevents) is not affected by measurement and is not contextual. It is in this sense that we can call the formulation \emph{realist}. Furthermore, it was shown that there exists a unique classical domain (unlike consistent histories), and that probabilistic predictions can be recovered by the use of Cournot's principle. Other recent developments and future directions were stated in section \ref{4.4}.

\ack The author is grateful to Rafael Sorkin for introducing and developing the coevents formulation. He also wants to thank him and Fay Dowker, Sumati Surya and Yousef Ghazi-Tabatabai for many discussions. This work is partly supported by COST Action MP1006 ``Fundamental Problems in Quantum Physics''.

\end{document}